# Polar Dynamics at the Jahn-Teller Transition in Ferroelectric GaV$_4$S$_8$

Zhe Wang,[1] E. Ruff,[1] M. Schmidt,[1] V. Tsurkan,[1,2] I. Kézsmárki,[3,4] P. Lunkenheimer,[1,*] and A. Loidl[1]

[1]*Experimental Physics V, Center for Electronic Correlations and Magnetism, Institute of Physics, University of Augsburg, 86135 Augsburg, Germany*
[2]*Institute of Applied Physics, Academy of Sciences of Moldova, Chisinau MD-2028, Republic of Moldova*
[3]*Department of Physics, Budapest University of Technology and Economics*
[4]*MTA-BME Lendület Magneto-optical Spectroscopy Research Group, 1111 Budapest, Hungary*

We present a dielectric spectroscopy study of the polar dynamics linked to the orbitally driven ferroelectric transition in the skyrmion host GaV$_4$S$_8$. By combining THz and MHz-GHz spectroscopy techniques, we succeed in detecting the relaxational dynamics arising from coupled orbital and polar fluctuations in this material and traced its temperature dependence in the paraelectric as well as in the ferroelectric phase. The relaxation time significantly increases when approaching the critical temperature from both sides of the transition. It is natural to assume that these polar fluctuations map the orbital dynamics at the Jahn-Teller transition. Due to the first-order character of the orbital-ordering transition, the relaxation time shows an enormous jump of about five orders of magnitude at the polar and structural phase transition.

PACS numbers: 77.84.-s, 71.70.Ej, 75.85.+t, 75.25.Dk

Ferroelectricity plays an important role in many areas of modern technology, e.g., for opto-electronic devices, piezo actuators, non-volatile computer memories, or in capacitor engineering for future energy-storage devices. While the polarization in classical ferroelectrics arises from the structurally driven displacement of ions, in recent years alternative routes to ferroelectric (FE) order have come into the focus of interest. This includes the so-called improper, spin-driven ferroelectricity found in many multiferroic materials combining polar and magnetic order [1,2] and also electronic ferroelectricity, where the FE order primarily arises from electronic degrees of freedom without major contributions from ionic displacements [3,4,5,6]. Another intriguing discovery is orbital-order driven ferroelectricity which, triggered by an intensive research for new functionalities in transition-metal compounds, has been identified in a number of materials [7,8,9,10,11].

In many compounds and especially at high temperatures, depending on the electronic configuration, the spatial orientations of electronic orbitals are degenerate with equal probability densities along symmetry-equivalent crystallographic directions. In compounds with partially filled *d* shells, long-range orbital order can be established via the cooperative Jahn-Teller effect, driven by the instability of orbital degeneracy with respect to elastic distortions [12]. Orbital order can also arise via exchange interactions and spin-lattice coupling [13,14,15]. Orbital order and ferroelectricity are usually decoupled from each other and only few FE Jahn-Teller systems are known until now [7,8,9,10]. Very recently it has been found that FE polarization originating from a cooperative Jahn-Teller distortion also shows up in some lacunar spinels, including GaV$_4$S$_8$ investigated in the present work [11,16].

Canonical FE compounds are well established and usually classified into displacive and order-disorder systems, despite the fact that in the majority of real materials a strict classification scheme is hardly appropriate [17,18]. Within a somewhat oversimplified but instructive picture, the polar moments in the former systems only emerge below the transition, while in the latter they already exist at high temperatures and order to produce a net polarization below the phase-transition. A more rigorous description of displacive ferroelectrics is achieved in terms of soft mode dynamics within a single-well potential, while in order-disorder systems molecular rearrangements, e.g., of electric dipoles or hydrogen bonds, are treated in multi-well potentials [17,18]. Perovskite-derived ferroelectrics and hydrogen-bonded KH$_2$PO$_4$ are prototypical examples for these two limiting scenarios. In the former systems, the frequency of a specific phonon mode goes to zero, as often experimentally detected [18,19] and the displacement pattern condenses into a new equilibrium position. In the latter, soft collective excitations of local electric dipoles give rise to a dielectric response that is in most cases of relaxational type and usually detected at much lower frequencies (typically below GHz) [18,20,21,22,23,24,25].

In the system under consideration in this Letter, ferroelectricity was suggested to be driven by the orbital degrees of freedom [16]. As orbital fluctuations already exist above the Jahn-Teller transition, the ferroelectricity can be speculated to be of order-disorder type. Our aim is to check the presence and nature of the dipolar relaxation mechanism inducing ferroelectricity in this system, which should simultaneously reflect the entanglement between the polar and orbital dynamics. Coupled polar and orbital dynamics were detected earlier in the orbital-glass system FeCr$_2$S$_4$ by dielectric spectroscopy [26]. For this purpose we have performed a combined THz and radio/microwave-frequency (MHz-GHz) dielectric investigation. We find polar fluctuations in the THz regime above as well as in the low-frequency regime below the Jahn-Teller transition. This finding helps unravelling the nature of the FE phase transition in this system and of orbital-order driven ferroelectricity in general.

Lacunar spinels are built up by only weakly linked molecular clusters $(AX_4)^{n-}$ and $(M_4X_4)^{n+}$ with well-defined local electron densities and composite molecular spin states. [27,28]. The lacunar spinel $GaV_4S_8$ is a magnetic semiconductor with non-centrosymmetric symmetry at room temperature [29,30]. The cubane $(V_4S_4)^{5+}$ units form a face-centered cubic lattice and are separated by $(GaS_4)^{5-}$ tetrahedra. The $(V_4S_4)^{5+}$ units are characterized by two interpenetrating tetrahedra made up by the four vanadium atoms and four triply-bridging sulfur atoms [31]. These cubane clusters constitute one unique molecular electronic distribution characterized by a local spin $S = \frac{1}{2}$ of an unpaired electron occupying a triply degenerate orbital. The orbital degeneracy is lifted by a cubic-to-rhombohedral structural phase transition at $T_{JT} = 44$ K and magnetic ordering takes place at $T_m = 12.7$ K [29,30]. The structural phase transition has been characterized as a Jahn-Teller transition leading to an elongation of the $V_4$ tetrahedra along any of the four crystallographic <111> directions [32], establishing a multi-domain rhombohedral state. Recently it has been found that $GaV_4S_8$ exhibits a complex low-temperature magnetic phase diagram, with a cycloidal and a skyrmion phase embedded within a collinear ferromagnetic phase [33]. The skyrmion lattice consists of Néel-type skyrmions and appears just below $T_m$ at low external magnetic fields. In addition, sizable FE polarization of ~ 1 μC/cm$^2$ appears below the Jahn-Teller transition and spin-driven excess polarizations appear in all magnetic phases, making $GaV_4S_8$ a unique multiferroic compound exhibiting orbital-induced ferroelectricity and a skyrmion-lattice phase with spin vortices dressed with electric polarization [16].

The preparation and characterization of high quality $GaV_4S_8$ samples has been described in Ref. [33]. Measurements in the radio-frequency and microwave range (1 MHz < $\nu$ < 3 GHz) at temperatures from 4 to 300 K were performed using a reflectometric technique with the sample capacitor mounted at the end of a coaxial line [34]. For these measurements an Agilent E4991A impedance analyzer was employed and we used platelet-shaped single crystals of size 1 mm$^2$ and 0.25 mm thickness, with the large (111) surface of these samples contacted with silver paste. THz experiments in transmission [35], perpendicular to the (111) plane, were carried out with a TeraView time-domain spectrometer for frequencies from 300 GHz to 3 THz and temperatures from 4 K up to room temperature. In these experiments, transmission and phase shift are simultaneously measured and, hence, the complex permittivity can directly be deduced.

Figure 1 shows the temperature dependence of the complex dielectric permittivity $\varepsilon^* = \varepsilon' - i\varepsilon''$ for various frequencies between 540 GHz and 2.5 THz as determined via THz spectroscopy between room temperature and 4 K. Above the Jahn-Teller transition, the dielectric constant $\varepsilon'$ [Fig. 1(a)] shows significant frequency dependence: For the lowest frequency, a Curie-Weiss like increase is observed. At higher frequencies it is followed by a continuous decrease at lower temperatures yielding a broad maximum, which is located around 200 K at the highest frequencies. This maximum shifts to lower temperatures with decreasing frequency. The peak becomes finally shifted below the Jahn-Teller transition temperature $T_{JT} = 44$ K for frequencies < 0.9 THz. However, below $T_{JT}$ all these strong frequency-dependent effects become abruptly suppressed and $\varepsilon'$ only exhibits weak frequency and temperature dependence. Similar findings become apparent in the temperature dependence of the dielectric loss $\varepsilon''$ in Fig. 1(b): A peak maximum, which appears just below 100 K at the highest frequency, rapidly shifts below $T_{JT}$ for lower frequencies, leaving only the high-temperature increase observable. Again this dispersion scenario becomes drastically suppressed below the orbital-ordering transition.

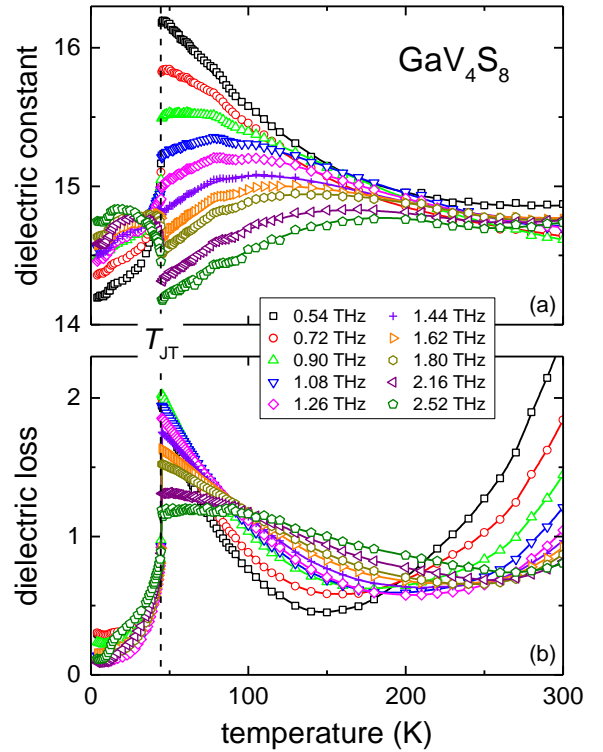

FIG. 1 (color online). Temperature dependence of dielectric constant (a) and dielectric loss (b) in $GaV_4S_8$ for a series of frequencies between 0.5 and 2.5 THz. The dashed vertical line indicates the Jahn-Teller transition at $T_{JT} = 44$ K.

Notably, the dielectric constant shows small but significant anomalies in the region below $T_{JT}$ when crossing the magnetic phase boundary at $T_m = 12.7$ K pointing to magnetoelectric effects as indeed reported in Ref. [16]. The increase of the dielectric loss towards high temperatures stems from conductivity contributions. The observed features in the temperature-dependent permittivity at $T > T_{JT}$, shifting to lower temperature with decreasing frequency, point to relaxational behavior as expected for ferroelectrics of order-disorder type. However, in contrast to canonical order-disorder ferroelectrics [18,21,22,23,24,25], the frequency-dependent peaks in $\varepsilon'(T)$ and $\varepsilon''(T)$ of GaV$_4$S$_8$ are truncated at temperatures below the phase transition. It seems reasonable to ascribe this finding to the first-order character of the transition [16].

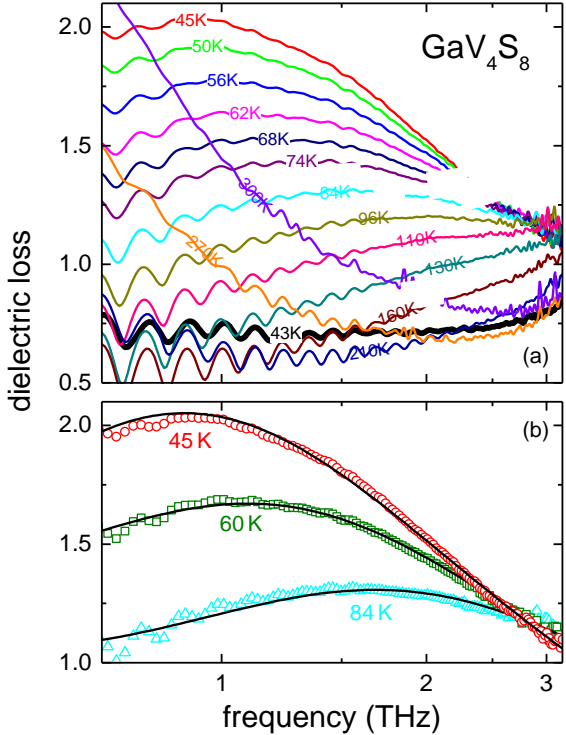

FIG. 2 (color online). (a) Frequency dependence of the dielectric loss between 0.5 and 3 THz in GaV$_4$S$_8$. Data are shown in a semilogarithmic plot for a series of selected temperatures between room temperature and 43 K (bold line), just below the Jahn-Teller transition at 44 K. (b) Dielectric loss as measured at three temperatures (open symbols) and fits with Eq. (1) using $\alpha = 0$ (solid lines).

To get more quantitative information about this relaxation process, Fig. 2(a) shows the dielectric loss $\varepsilon''$ as function of frequency for a series of temperatures from 300 K to just below the Jahn-Teller transition. At the highest temperatures (300 and 270 K) the increase of $\varepsilon''$ towards low frequencies signals contributions from dc conductivity, which leads to a low-frequency divergence via the relation $\varepsilon''_{dc} \propto \sigma_{dc}/\nu$ and becomes rapidly suppressed on decreasing temperatures. At lower temperatures, a loss peak indicating a relaxational process shifts into the experimental frequency window. It moves towards lower frequencies with an amplitude gradually increasing with decreasing temperature. At 45 K, just 1 K above the Jahn-Teller transition, the peak maximum is close to 1 THz. However, the relaxation peak becomes abruptly suppressed below the transition as documented by the low and frequency-independent loss observed at 43 K, just 1 K below the phase transition. In the orbitally ordered phase the loss is low and close to experimental uncertainty [cf. Fig. 1(b)] implying that the relaxation process has shifted out of the frequency window.

The frequency dependent permittivity data obtained in the THz range are well reproduced by a purely relaxational behavior plus charge-transport contribution:

$$\varepsilon^* = -i\sigma_{dc}/\varepsilon_o\omega + \Delta\varepsilon/[1 + (i\omega\tau)^{1-\alpha}] \qquad (1)$$

The first and second terms of Eq. (1) describe the contributions from dc conductivity and the dipolar relaxation, respectively. Here $\sigma_{dc}$ is the dc conductivity, which is important at high temperatures, $\varepsilon_0$ is the permittivity of free space, and $\Delta\varepsilon$ is the relaxation strength. $\tau$ corresponds to the mean relaxation time of the relaxing entities, which in our case are simultaneous structural, polarization, and orbital fluctuations which are highly entangled in the present material. The parameter $\alpha$, which can vary between 0 and 1, accounts for a possible symmetric broadening of the relaxation peak and is zero in case of a pure Debye relaxation [36]. For clarity reasons, the fits to the dielectric loss are shown in Fig. 2(b) for three representative temperatures only. A single and monodispersive Debye relaxation with $\alpha = 0$ in Eq. (1) describes the frequency dependence reasonably well.

We found no trace of any relaxational response at $T > T_{JT}$ in the low-frequency range below 3 GHz, where polar fluctuations of order-disorder ferroelectrics are often reported [17,18,21,22,23,24,25]. Thus, we conclude that the detected relaxation process evidenced in Figs. 1 and 2 indeed reflects the FE fluctuations, which are caused by orbital reorientations above the Jahn-Teller transition coupled to dipole-active lattice distortions. At the heart of this mechanism lies the elongation of the tetrahedral V$_4$ units along the crystallographic <111> direction, induced by orbital order [16].

Figures 1 and 2 provide experimental evidence for the slowing down of dipolar relaxations and a concomitant increase of the mean relaxation time when approaching the Jahn-Teller transition from above. As for $T < T_{JT}$ this relaxation obviously is not located in the THz frequency range, it seems possible that it abruptly shifts to lower frequencies when passing the structural phase transition. To address this issue, we measured dielectric spectra in the FE state below $T_{JT}$, covering the radio- to microwave frequency range. Figure 3 shows a representative set of the results. Here the loss is plotted vs. frequencies from 1 MHz to 3 GHz for a series of temperatures below $T_{JT}$. Indeed, the relaxation peak appears in this frequency regime and shifts towards higher frequencies with decreasing temperatures as expected below the polar phase transition [17,18]. Again the peaks can be fitted assuming relaxational behavior as given by Eq. (1) (solid lines in Fig. 3). The fits work rather well if slight deviations at the highest frequencies are neglected. The obtained width parameter $\alpha$ deviates from zero and is about 0.38 with very weak temperature dependence. This indicates polydispersive behavior corresponding to a distribution of relaxation times, which can be explained by heterogeneities in the multi-domain state of the ordered polar phase and/or by domain-oscillation effects. Dc conductivity contributions are still present just below the structural phase transition but become almost negligible at the lowest temperatures.

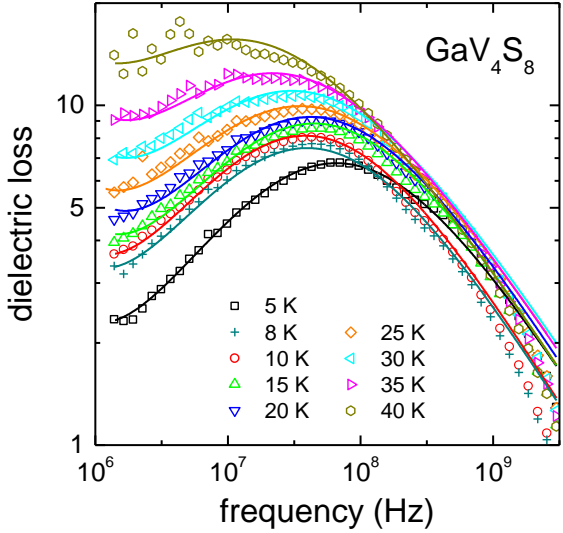

FIG. 3 (color online). Dielectric loss spectra of $GaV_4S_8$ measured in the MHz-GHz range for a series of temperatures between 5 and 40 K, shown in double-logarithmic representation (symbols). The solid lines represent results of fits with Eq. (1).

The temperature dependences of the relaxation time resulting from the fits above and below the Jahn-Teller transition are provided in Fig 4(a). We observe a clear increase of $\tau(T)$ when approaching the phase-transition temperature from above and below, signaling a slowing down of the critical dynamics as often observed at FE phase transitions [18,21,22,23,24,25]. At $T_{JT}$ the mean relaxation time abruptly jumps from the sub-picosecond time scale to values well above nanoseconds. What is the reason for this strong jump? Coming from high temperatures, $\tau(T)$ starts increasing. For the typical second-order transitions of order-disorder ferroelectrics, it approaches high values above nanoseconds when coming close to the transition temperature before decreasing again below the transition [21,22,23,24,25]. However, the Jahn-Teller phase transition in $GaV_4S_8$ is of first order [16]. Thus, before $\tau(T)$ can reach high values, the first-order transition intervenes and, consequently, the relaxational dynamics for $T > T_{JT}$ is always fast with only a moderate slowing down.

In the high-temperature cubic phase, $\tau(T)$ close to the phase transition can be described by a critical law $\tau \propto 1/(T - T_c)^\beta$ [solid line in Fig. 4(c)], with an exponent $\beta = 1$ as observed in order-disorder ferroelectrics and predicted by mean-field theory [17,18,21,22,24,25]. The critical temperature is found to be $T_c \sim 14$ K, located well below the orbital-ordering phase transition at 44 K [Fig. 4(c)]. This indicates that at the Jahn-Teller transition the polar dynamics is far from a divergence of the mean relaxation time. Indeed, one has to be aware that the phase transition is of first order character and an analysis in terms of critical laws is questionable. Below $T_{JT}$, within the FE phase, the relaxation time can be described by a critical law, too, but with an exponent $\beta = 0.5$ [Fig. 4(b)]. Here the dynamic properties should be dominated by domain motions as present in all ferroelectrics and, hence, the significance of this exponent parameter remains unclear. Notably, in the low-temperature phase the critical temperature is close to $T_{JT}$. Interestingly, the magnetic phase transition at 12.7 K seems to have a clear effect on the temperature dependence of the relaxation time, leading to a step-like decrease in Fig. 4(b) (i.e., an increase of $\tau$). In the magnetically ordered phase the relaxation time decreases significantly towards the lowest temperature [Fig. 4(b)].

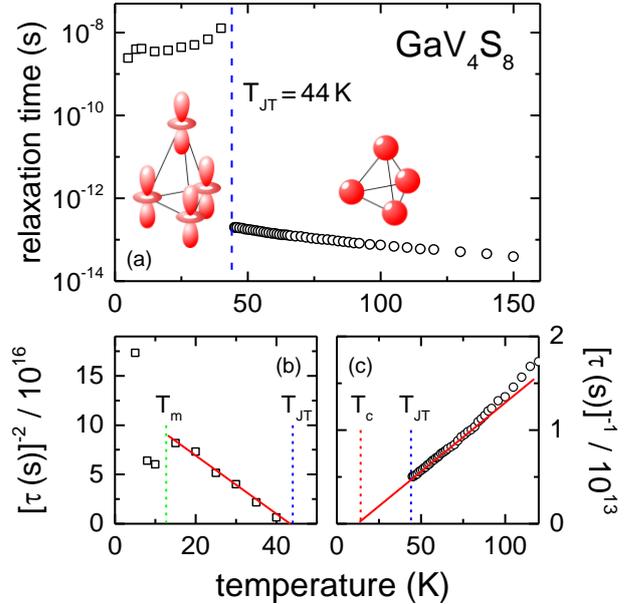

FIG. 4 (color online). Temperature dependence of the mean relaxation time in $GaV_4S_8$. (a) Relaxation times above and below the Jahn-Teller transition with a jump by 5 orders of magnitude at the first-order phase transition. The distortion of the $V_4$ tetrahedron by orbital order is schematically indicated in the inset. (b) and (c): Critical scaling of the mean relaxation time below and above the orbital-order driven FE phase transition, respectively. The solid lines represent linear fits. The vertical dashed lines indicated the different characteristic temperatures of the system (see text).

In this work, by utilizing THz and radio- to microwave spectroscopy we have found clear experimental evidence for polar relaxational dynamics associated with an orbital-order driven FE transition. Similar to conventional order-disorder ferroelectrics, this dynamics exhibits critical slowing down both above and below the transition, however, with highly non-canonical behavior at the transition. In $GaV_4S_8$, the $V_4$ molecular units are strongly distorted from their ideal high-temperature tetrahedral shape, as indicated in the inset of Fig. 4(a). This distortion creates a shift of ions, a redistribution of electron density, and long-range polar order [16]. We propose that the observed critical dynamics is driven by orbital fluctuations and measures the time scale of orbital reorientations in a multi-well potential, closely coupled to polar degrees of freedom. As the Jahn-Teller transition in $GaV_4S_8$ is of first order, above $T_{JT}$ the time scales characterizing the fluctuations never reach high values but abruptly slow down by about five orders of magnitude when entering the FE and orbitally ordered state.

This research was supported by the Deutsche Forschungsgemeinschaft (DFG) via the collaborative research center TRR 80 "From Electronic Correlations to Functionality" (Augsburg, Munich, Stuttgart) and by the Hungarian Research Fund OTKA K 108918.

* Corresponding author.
Peter.Lunkenheimer@Physik.Uni-Augsburg.de